\documentclass[aps,prb,showpacs,manuscript]{revtex4}
\usepackage{graphicx}
\begin{document}
\title{Relation between Effective Conductivity and  Susceptibility 
of Two -- Component Rhombic Checkerboard}

\author{Leonid G. Fel}
\affiliation{School of Physics and Astronomy of Exact Sciences
Tel Aviv University, Ramat Aviv 69978, Israel}
\author{Ilia V.  Kaganov}
\affiliation{Department of Chemical Engineering, Technion-IIT,
Haifa 32000, Israel}
\date{\today}

\begin{abstract}
The heterogeneity of composite leads to the extra charge concentration at the
boundaries of different phases that results essentially nonzero effective
electric susceptibility. 
The relation between tensors of effective electric susceptibility 
$\widehat\chi_{ef}$ and effective conductivity $\widehat\sigma_{ef}$
of the infinite two--dimensional two--component regular composite with rhombic 
cells structure has been established. The degrees of electric field singularity at 
corner points of cells are found by constructing the integral equation for the 
effective conductivity problem. The limits of weak and strong 
contrast of partial conductivities $\sigma_1,\sigma_2$ are considered. 
The results are valid for thin films and cylindrical samples.
\end{abstract}
\pacs{73.25.+i, 73.40.-c, 73.40.Jn, 73.50.-h, 73.61.-r}
\keywords{2D two component composite, effective conductivity
and susceptibility,
electric field singularity}
\maketitle

\section{Introduction}
The evaluation of effective properties for two--dimensional (2D) two--component 
composites, which determine the behavior of the medium at large scales, 
given rise by Keller \cite{keller64} and  Dykhne \cite{dykhne70}, 
remains a topic of high activity. Among different approaches (variational bounds 
\cite{ber78,mil81}; asymptotic \cite{kell87,kozlov89}; numerical 
\cite{hels91}; network analogue \cite{luck91,khfel01}) used to consider 
this problem, the analytical approach, being classical problem of mathematical 
physics, is surprisingly very difficult. Exact values of effective parameters 
are of great interest even though these values are established in idealized 
models. It seems that explicit formulae are available only as exceptions.
Such formulae which solve the field equations were obtained for two--component 
regular checkerboard with square \cite{berd85}, rectangular \cite{obnos96,milt2001} 
and triangular \cite{obnos99} unit cell using complex--variables analysis.  
Another technique (integral equations)
was used in the recent papers dealt with square \cite{ovch00} and triangular  
\cite{ovch02} regular checkerboard. 

Almost all these studies were directed towards effective conductivity $\sigma_{ef}$ 
evaluation despite of important fact that the heterogeneity contributed to the 
conducting composite some dielectric properties. The homogeneous metal does not possess 
static dielectric properties (such as electric susceptibility) because only core 
electrons can contribute there, but their influence is obviously small. However, 
heterogeneity leads to the extra charge concentration at the boundaries of 
different phases, which results essentially nonzero effective electric 
susceptibility $\chi_{ef}$. The implication follows that the relation between the 
conductivity $\widehat\sigma_{ef}$ and susceptibility $\widehat\chi_{ef}$ 
effective tensors must exist. 

In the present paper we will consider the regular 2D two--component rhombic checkerboard
and derive such relation. This middle--symmetric structure belongs to 
$p^{\prime}_{c}mm\;$--plane group \cite{belov56} and gives rise to anisotropy 
of $\widehat\sigma_{ef}$. In some sense this anisotropic model is more 
universal than the regular 2D two--component rectangular checkerboard 
($c^{\prime}mm\;$--plane group). Really, the effective electric properties are mostly 
determined by the corner points of the cell, where the electric field is singular
\cite{kell87,kozlov89}. The structure of composite near these points in rhombic 
checkerboard is governed by {\it arbitrary} angular variable.

\section{Integral Equation}
\begin{figure}[ht]
\centerline{\includegraphics*[scale=0.5]{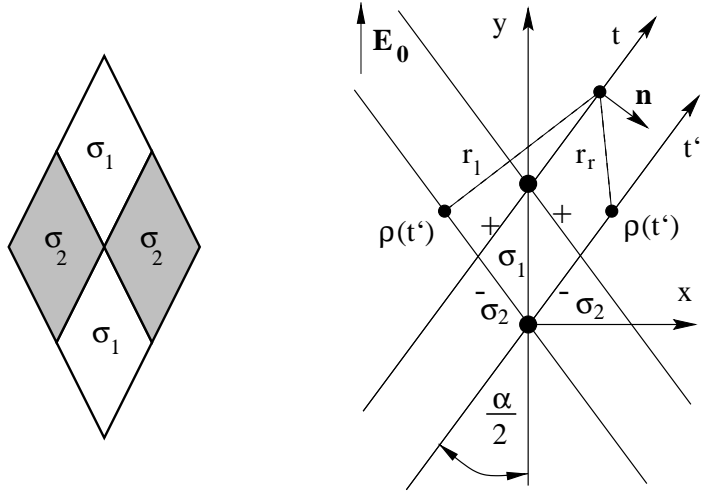}}
\caption{Regular rhombic two -- component checkerboard under electric field $E_0$:
unit cell ({\sl left}) and  basic variables for integral equation ({\sl right}). 
The distribution of the charges is drawn in accordance with chosen inequality 
$\sigma_1\geq \sigma_2$.}
\label{rom1}
\end{figure}

The regular checkerboard structure is composed of rhombic
conducting cells with isotropic homogeneous conductivities $\sigma_1$ and
$\sigma_2$, hereafter $\sigma_1\geq \sigma_2$. The backbone of such structure can be
represented as the set of images of the letter "{\sf X}" with infinitely long 
legs, which are shifted up and down to distance $2 N \cos{\alpha\over 2}$,
$N=0,\pm1,\pm2,\ldots$; $\alpha$ is the smallest angle between legs 
(see Figure \ref{rom1}). The side of the cell is scaled by unit length. Such 
arrangement of the checkerboard allows to generate the kernel for the integral 
equation by single series summation rather than by double summation \cite{ovch00,ovch02}.

We will consider the external unit field $E_0$ be applied in vertical direction $Y$. 
It is one of the principal axes of the effective material tensors.  Another axis 
$X$ may be considered just changing $\alpha\to\pi -\alpha$.

Let us proceed with solution of Laplace equation for a scalar potential $\phi(r)$ 
at the infinite plane $S$
\begin{eqnarray} 
\phi({\bf r})=-E_0 y - 4 \pi \int_S G({\bf r},{\bf r_1}) 
\rho({\bf r_1})\;d^2 r_1\;,\;\;\;G({\bf r},{\bf r_1})=
\frac{1}{2\pi} \ln |{\bf r}-{\bf r_1}|\;,
\label{int1}  
\end{eqnarray}
where $G({\bf r},{\bf r_1})$ is the two-dimensional 
Green function and $\rho({\bf r})$ is a charge distribution at the plane. The 
boundary conditions at the edge relate normal components of the field $E_n$ and of 
the current density $j_n$ 
\begin{eqnarray}
E_n^{(1)}-E_n^{(2)}=4\pi \rho(t)\;,\;\;\;j_n(t)=\sigma_1E_n^{(1)}=\sigma_2E_n^{(2)}\;,
\label{int3}
\end{eqnarray}
where a new variable $t$ is introduced to measure the distance along the edge of a 
unit cell counted from the cell corner and $\rho (t)$ hereafter is the 
charge distribution at the edge. 
The boundary conditions (\ref{int3}) allow to write master equation 
\begin{eqnarray}
E_n^{(1)}+E_n^{(2)}=\frac{4\;\pi}{Z} \rho(r)\;,\;\;\;
Z=\frac{\sigma_1-\sigma_2}{\sigma_1+\sigma_2}\;,\;\;\;0\leq Z\leq 1\;.
\label{int6}
\end{eqnarray}
Finding the corresponding derivatives $E_n^{(i)}$ (see Appendix \ref{appendix1}) 
we come to the integral equation
\begin{eqnarray}
\frac{2\;\pi\;g(t)}{Z} \rho(t)=\sin \frac{\alpha}{2}-
4 \int_{-\infty}^{+\infty} \rho(t^{\prime})K_1(t,t^{\prime})\;d t^{\prime}\;,
\label{int7}
\end{eqnarray}
where a new function $g(t)$ reflects a periodic interchange of the constituents 
($\sigma_1$ and $\sigma_2$) with variation of argument $t$
\begin{eqnarray}
g(t)=\mbox{sgn[mod}(t,2)-1]\;,\;\;g(-t)=-g(t)\;;\;\;g(t)=-1\;\;\;\mbox{for}\;\;\;
0< t\leq 1\;,
\label{int7c}
\end{eqnarray}
the function ${\rm mod}(t,2)$ gives the remainder on division of $t$ by 2 and 
${\rm sgn}[x]$ 
gives -1, 0 or 1 depending on whether $x$ is negative, zero, or positive. The kernel 
$K_1(t,t^{\prime})$ is given by formula
\begin{eqnarray}
K_1(t,t^{\prime})=\sum_{k=-\infty}^{+\infty}
\left[\frac{k}{(t-t^{\prime})^2 \tan \frac{\alpha}{2}+
(t-t^{\prime}-2k)^2 \cot \frac{\alpha}{2}}+
\frac{k+t^{\prime}}{(t+t^{\prime})^2 \tan \frac{\alpha}{2}+
(t-t^{\prime}-2k)^2 \cot \frac{\alpha}{2}}\right]
\label{int7a}
\end{eqnarray}
It is worth to represent the kernel for further summation as
\begin{eqnarray}
K_1(t,t^{\prime})=\frac{1}{4}\tan \frac{\alpha}{2}
\sum_{k=-\infty}^{+\infty}\left[\frac{k}{(k-k_1)(k-k_1^*)}+   
\frac{k+t^{\prime}}{(k-k_2)(k-k_2^*)}\right]\;,\label{int7b}
\end{eqnarray}
where the zeros $k_1,k_1^*$ and $k_2,k_2^*$ of both denominators in (\ref{int7a}) read 
\begin{eqnarray}
k_1,\;k_1^*=\frac{t-t^{\prime}}{2}\pm i\;
\frac{|t-t^{\prime}|}{2}\tan \frac{\alpha}{2}\;,
\;\;\;\;
k_2,\;k_2^*=\frac{t-t^{\prime}}{2}\pm i\;
\frac{|t+t^{\prime}|}{2}\tan \frac{\alpha}{2}\;.
\label{int8}
\end{eqnarray}
Making use of identity 
\begin{eqnarray}
\frac{k+t^{\prime}}{(k-k_i)(k-k_i^*)}=
{\rm Im}\left(\frac{k_i+t^{\prime}}{k-k_i}\right)\frac{1}{{\rm Im}\;k_i}\;,
\nonumber
\end{eqnarray}
we can evaluate (\ref{int7b}) in the sense of principal value and 
reduce essentially the kernel of integral equation (\ref{int7})
\begin{eqnarray}
K_1(t,t^{\prime})=-\frac{\pi}{2}{\rm Im}\left\{
\frac{k_1\cdot \cot \pi k_1}{|t-t^{\prime}|}+
\frac{(k_2+t^{\prime})\cdot \cot \pi k_2}{|t+t^{\prime}|}\right\}\;.
\label{int11}
\end{eqnarray}
The further simplification of the kernel $K_1(t,t^{\prime})$ can be continued 
by usage of trigonometry. 
Introducing $\widetilde\rho (t)=\rho (t) g(t)$ we obtain finally the integral
equation
\begin{eqnarray}
-\frac{2}{Z}\widetilde\rho(t)=-\frac{1}{\pi} \sin \frac{\alpha}{2}+
\int_{-\infty}^{+\infty}
\widetilde\rho(t^{\prime}) K_2(t,t^{\prime})g(t^{\prime})\;d t^{\prime}\;,
\label{int12}
\end{eqnarray}
where 
\begin{eqnarray}
K_2(t,t^{\prime})=
\frac{\tan \frac{\alpha}{2} \sin \pi (t-t^{\prime})-
\sinh \left(\pi (t-t^{\prime})\tan \frac{\alpha}{2}\right)}
{\cos \pi (t-t^{\prime})-\cosh \left(\pi (t-t^{\prime})\tan 
\frac{\alpha}{2}\right)}+
\frac{\tan \frac{\alpha}{2} \sin \pi (t-t^{\prime})-
\sinh \left(\pi (t+t^{\prime})\tan \frac{\alpha}{2}\right)}
{\cos \pi (t-t^{\prime})-\cosh \left(\pi (t+t^{\prime})\tan 
\frac{\alpha}{2}\right)}
\nonumber
\end{eqnarray}
The function $\rho(t)$ being a solution of integral equation (\ref{int12}) makes 
it possible to find an exact expression of the effective conductivity tensor
$\widehat\sigma_{ef}$ (see Section \ref{suscept}).
\section{Asymptotic behavior of $\rho(t)$ near the corners}
\label{corner}
We start  this Section with two algebraic properties of the function  $\rho(t)$,  
which will be used in order to simplify further calculation. These are the 
parity and periodicity of the functions $\widetilde\rho(t)$, $\rho(t)$,
which are following from (\ref{int12}) 
\begin{eqnarray}
\widetilde\rho(-t)=\widetilde\rho(t)\;,\;\;\;
\widetilde\rho(t+2)=\widetilde\rho(t)\;\;\longrightarrow\;\;
\rho(-t)=-\rho(t)\;,\;\;\;
\rho(t+2)=\rho(t)\;.
\label{int12a}
\end{eqnarray}
They are in full agreement with physics of the charge distribution $\rho(t)$ 
along the edges of the cells. A proof follows from
an accurate evaluation of integral in (\ref{int12}). Indeed, the parity property 
follows due to (\ref{int7c}) and identity $K_2(-t,-t^{\prime})=-K_2(t,t^{\prime})$
\begin{equation}
\frac{2}{Z}\widetilde\rho(-t)+\frac{1}{\pi} \sin \frac{\alpha}{2}=
-\int_{+\infty}^{-\infty}
\widetilde\rho(-t^{\prime}) K_2(-t,-t^{\prime})g(-t^{\prime})\;d t^{\prime}=
\int_{-\infty}^{+\infty} 
\widetilde\rho(-t^{\prime}) K_2(t,t^{\prime})g(t^{\prime})\;d t^{\prime}\;.
\end{equation}
The periodicity could be proven in the similar way.

A similar integral equation was appeared in
Ref. \onlinecite{ovch00} for the two -- component
checkerboard with square unit cell. Its solution is presented by the 
means of Weierstrass elliptic function $\rho_{sq}(t)\propto {\bf \wp}^{\kappa}(t)$, 
where $\sin \pi \kappa=Z$, and is found by inspection its behavior near the 
branch points $t=0$ and $t=1$
\begin{eqnarray}
E_n^{(i)}(t)\sim
\rho_{sq}(t) \stackrel{t\rightarrow 0}
\sim \frac{1}{t^{2\lambda_0}}\;\;,\;\;\;
E_n^{(i)}(t)\sim \rho_{sq}(t) \stackrel{t\rightarrow 1}
\sim (1-t)^{2\lambda_1}\;\;,\;\;\;\lambda_0=\lambda_1=\kappa\;.
\label{int13}
\end{eqnarray}
The equality of the exponents $\lambda_0=\lambda_1$ is here essential. Already the
two--component checkerboard with triangle unit cell \cite{ovch02} breaks the validity 
of (\ref{int13}), that made unattainable an explicit 
\footnote{The explicit solution of effective conductivity problem for 
two--component checkerboard, composed of the perfect triangles, was obtained 
in Ref. \onlinecite{obnos99} where the conformal mapping of the triangle on the unit 
circle with a cut was used.}
solution, but only an efficient approximate method was proposed.

The rhombic structure, discussed in the present paper, also gives rise to distinct  
exponents. Asymptotic behavior of $\rho(t)$ near the branch points $t=0$ and 
$t=1$ can be found from the equation (\ref{int12}) (see Appendix \ref{appendix1})
\begin{eqnarray}
E_n^{(i)}(t)\sim \rho(t)\stackrel{t\rightarrow 0}
\sim  \frac{1}{t^{2\mu_0}}\;&,&\;
\sin \pi \mu_0=
Z\sin\left[\alpha+\mu_0\left(\pi-2\alpha\right)\right]\;,\label{int14}\\
E_n^{(i)}(t)\sim \rho(t)\stackrel{t\rightarrow 1}
\sim  (1-t)^{2\mu_1}\;&,&\;
\sin \pi \mu_1=
Z\sin\left[\alpha+\mu_1\left(2\alpha-\pi\right)\right]\;.
\label{int15}
\end{eqnarray}
Then
\begin{eqnarray}
\mu_0(Z)=\mu_1(Z)=\kappa=\frac{1}{\pi}\arcsin Z\;\;\;
\mbox{if}\;\;\;\alpha=\frac{\pi}{2}\;;\;\;\;\;
\mu_0(Z)=-\mu_1(-Z)\;\;\;\mbox{if}\;\;\;\alpha \neq \frac{\pi}{2}\;
\label{int16}
\end{eqnarray}
and for a large contrast in conductivities $\sigma_2\ll \sigma_1,\;Z\sim 1$: $
2\mu_0=1,\;2\mu_1=\alpha/(\pi-\alpha)$.

This shows that the generic rhombic cell ($\alpha \neq \frac{\pi}{2}$)
does not lead to the solution of (\ref{int12}), which can be built out by 
simple rescaling of Weierstrass elliptic functions. An explicit solution of
integral equation remains to be performed. 

It turns out that the integral equation (\ref{int12}), obtained in the present  
Section, is sufficient to establish an exact relation between effective electric
conductivity $\widehat\sigma_{ef}$ and effective electric susceptibility 
$\widehat\chi_{ef}$, which was not to our knowledge discussed earlier.

\section{Effective Susceptibility of Rhombic Checkerboard}
\label{suscept}
Let us consider the  polarization of the rhombic checkerboard at the scales 
large, compared to the size of the cells. 
The effective electric susceptibility is the tensor which defined by ${\bf P}=
\widehat\chi_{ef} {\bf E}$ where ${\bf P}$ is polarization and ${\bf E}$ is
external electric field. In the reference frame (Figure \ref{rom1}), when 
$\widehat\chi_{ef}$ is diagonalized, its $y$--component is determined by 
induced dipole moment $d_y$ per unit square: $\chi_{ef}^y=d_y/S$, where
\begin{eqnarray}
d_y=\sum_{{\rm over\ all\atop charges\ q_j}}y_j q_j=
2\cos{\alpha\over 2}\sum_{k=-\infty}^\infty \int_{-\infty}^\infty (t+2k)\rho (t)dt
\label{int17}
\end{eqnarray}
is dipole moment of area $S=L_xL_y$ and $L_x,L_y$ are the sizes of a sample. The 
summation covers all induced edge--charges $q_j$ which are placed within the area $S$. 
The sample which is composed of $2N_x\times 2N_y$ unit cells has the area 
$S=2N_x \;2\sin\frac{\alpha}{2}\times 2N_y \;2\cos\frac{\alpha}{2}=
8N_x N_y\sin \alpha$. 
The formula (\ref{int17}) can be reduced making use of parity and periodicity
properties (\ref{int12a}) of $\rho(t)$. Its accurate evaluation reads
\begin{eqnarray}
d_y&=&4N_y\cos\frac{\alpha}{2}\int_{-\infty}^\infty t\rho (t)dt=
4N_y\cos\frac{\alpha}{2}\sum_{n=-N_x}^{n=N_x}\int_{2n-1}^{2n+1}t\rho (t)dt=
8N_xN_y\cos\frac{\alpha}{2}\int_{-1}^{1}t\rho (t)dt\;.\nonumber
\end{eqnarray}
Taking into account the inparity (\ref{int7c}) of the function $g(t)$ we obtain 
finally
\begin{eqnarray}
d_y=16 N_x N_y \cos\frac{\alpha}{2}\int_0^1 t\widetilde \rho (t)dt\;,\;\;\;\;
\mbox{and}\;\;\;\;\;\;
\chi_{ef}^y={1\over\sin{\alpha\over 2}}\int_0^1 t\widetilde\rho (t)\;dt\;.
\label{int18}
\end{eqnarray}
We define also the effective conductivity $\sigma_{ef}$ as a ratio of the
current $J=\int j_n(t)\;d t$ through the $x$--cross-section of the checkerboard
per the unit length to the applied field $E_0=1$
\begin{eqnarray}
\sigma_{ef}^y=\frac{4 \pi}{\sin \frac{\alpha}{2}} 
\frac{\sigma_1 \sigma_2}{\sigma_1-\sigma_2} \int_0^1 \widetilde \rho(t)\;d 
t\;,\;\;\;\;\;\sigma_{ef}^x(\alpha)=\sigma_{ef}^y(\pi-\alpha)\;.   
\label{int5}
\end{eqnarray}
Due to Keller \cite{keller64} the principal values of the tensor
$\widehat\sigma_{ef}$ satisfy the duality relations
\begin{equation}
\sigma_{ef}^x(\alpha)\cdot \sigma_{ef}^y(\alpha)=\sigma_1\sigma_2\;.
\end{equation}
Relating now two physical quantities (\ref{int18}),  (\ref{int5}), we 
 make use of auxiliary integral equation, obtained by integrating the 
equation (\ref{int12}) (see Appendix \ref{appendix2})
\begin{eqnarray}
\frac{\sigma_2}{\sigma_1-\sigma_2}\int_0^1 \widetilde\rho (t)\;dt=
\frac{1}{4\pi} \sin \frac{\alpha}{2}-\int_0^1 t\;\widetilde\rho (t)\;dt\;.
\label{int19}
\end{eqnarray}
The last relation could be rewritten in new notations (\ref{int18}),  (\ref{int5})
\begin{eqnarray}
4\pi\chi_{ef}^y=1-\frac{\sigma_{ef}^y}{\sigma_1}\;,\;\;
\mbox{and similarly}\;\;\;\;
4\pi\chi_{ef}^x=1-\frac{\sigma_{ef}^x}{\sigma_1}\;.
\label{int20}
\end{eqnarray}
One can think that $\sigma_1$, which appeared in (\ref{int20}), breaks the 
universality of the formulae. Actually, the denominator contains the maximal 
value of partial conductivities $\sigma_{\sf max}=\max(\sigma_1,\sigma_2)$.

In fact, we have established the tensorial relation in any reference frame
\begin{eqnarray}
4\pi\widehat\chi_{ef}=\widehat I-\frac{1}{\sigma_{\sf max}}\;\widehat\sigma_{ef}\;,
\label{int20a}
\end{eqnarray}
where $\widehat I$ is an identity matrix. Formula (\ref{int20a}) results in 
the particular square--checkerboard case: 
$4\pi\chi_{ef}^x=4\pi\chi_{ef}^y=1-\sqrt{\sigma_2/\sigma_1}$.

Let us consider now two different cases of the weak and large contrast in
partial conductivities. 

1. $(\sigma_1-\sigma_2)/\sigma_1\ll 1$, or $Z\ll 1$:

In the first order on $Z$ the integral equation (\ref{int12}) gives according to 
definition (\ref{int5})
\begin{equation}
\rho(t)=-\frac{Z}{2\pi} \sin \frac{\alpha}{2}\;\;\;\longrightarrow\;\;
\sigma_{ef}^x=\sigma_{ef}^y=\sigma_1(1-Z)
\end{equation}
and therefore
\begin{eqnarray}
4\pi\chi_{ef}^x=4\pi\chi_{ef}^y=Z\;.
\label{int21}
\end{eqnarray}

2. $\sigma_2\ll \sigma_1$:

In this limit the corner points $t\rightarrow 0$ of the cell become important. 
Making use of the distribution (\ref{int14}) for the fields and charge 
$E_n^{(i)} \sim\widetilde\rho (t)\sim t^{-2\mu_0}$ one can find approximately
in leading terms
\begin{eqnarray}
\mu_0= \frac{1}{2}-\frac{1}{\sqrt{\alpha(\pi -\alpha)}}
\sqrt{\frac{\sigma_2}{\sigma_1}}\;\;\;\;\longrightarrow\;\;
\sigma_{ef}^y= A\;\sqrt{\sigma_1 \sigma_2}\;. 
\label{int22}
\end{eqnarray}
Actually, the coefficient was found in \cite{kozlov89} :
$A=\sqrt{\alpha/(\pi-\alpha)}\;\cot (\alpha/2)$. Formula (\ref{int22}) implies
as well
\begin{eqnarray}
4\pi\chi_{ef}^x=1-A\;\sqrt{\frac{\sigma_2}{\sigma_1}}\;,\;\;\;
4\pi\chi_{ef}^y=1-\frac{1}{A}\;\sqrt{\frac{\sigma_2}{\sigma_1}}\;.
\label{int23}
\end{eqnarray}
\section{Conclusion}
\label{conc}

1. We have derived the integral equation for the effective conductivity problem
for the regular 2D two--component rhombic checkerboard. An asymptotic
behavior of the electric field was investigated near the singular points 
$t=0$ and $t=1$. 

2. The heterogeneity of composite leads to the extra charge concentration at the 
boundaries of different phases that results essentially nonzero effective 
electric susceptibility. 
The exact relation (\ref{int20a}) between the two most 
important electrical properties, namely, effective conductivity 
$\widehat\sigma_{ef}$ and effective susceptibility $\widehat\chi_{ef}$, 
of rhombic composite was established. An absence of specific angular parameter
$\alpha$ in this formula make us possible to conjecture its validity 
for any anisotropic two--component structure. It is shown that the tensor of 
electrical susceptibility has surprisingly simple structure in both cases of large 
and small contrast in partial conductivities $\sigma_1,\sigma_2$.

3. The relation derived in present paper is definitely valid for cylindrical samples.
It is also valid for thin films due to the conducting nature of the constituents, 
which confine the electric field inside the conductor.

\begin{acknowledgements}
This research was supported in part by grants from the U.S. -- Israel
Binational Science Foundation, the Israel Science Foundation, the Tel Aviv
University Research Authority, Gileadi Fellowship program of the Ministry
of Absorption of the State of Israel (LGF), and Israel Council for Higher 
Education (IVK).
\end{acknowledgements}

\appendix
\renewcommand{\theequation}{\thesection\arabic{equation}}
\section{Derivation of integral equation (\ref{int7}). Behavior of its  
solution near the branch points.}
\label{appendix1}
\setcounter{equation}{0}
We define the variables
\begin{eqnarray}   
x=t\sin \frac{\alpha}{2}\;,\;y=t\cos \frac{\alpha}{2}\;,\;
x^{\prime}=\pm t^{\prime}\sin \frac{\alpha}{2}\;,\;\;\;
y^{\prime}=t^{\prime}\cos \frac{\alpha}{2}+2k\cos \frac{\alpha}{2}
\label{appen1}
\end{eqnarray}
and normal vector to the edge
\begin{equation}
{\bf n}=\left(\cos \frac{\alpha}{2},-\sin \frac{\alpha}{2}\right)\;.
\end{equation}
Here $k$ is an ordinal number of the "{\sf X}" image. 
Taking in mind the contributions from the both left ({\sl l}) and right ({\sl r}) 
edges of the rhombic tile we will find the derivative $\partial \phi/\partial n=
({\bf n}\;\nabla)\;\phi$
\begin{eqnarray}
-\frac{\partial \phi}{\partial n}=E_0 \sin \frac{\alpha}{2}+
\sum_{k=-\infty}^{\infty}\int_{-\infty}^{\infty} d t^{\prime}\rho(t^{\prime})
\frac{1}{r_r^2(t,t^{\prime})}
\left(\frac{\partial r_r^2(t,t^{\prime})}{\partial n}\right)+
\sum_{k=-\infty}^{\infty}\int_{-\infty}^{\infty} d t^{\prime}\rho(t^{\prime})
\frac{1}{r_l^2(t,t^{\prime})}
\left(\frac{\partial r_l^2(t,t^{\prime})}{\partial n}\right),
\nonumber
\end{eqnarray}
which lead after simple algebra to equation 
\begin{eqnarray}
\frac{1}{2}\left(E^{(1)}_n+E^{(2)}_n\right)=- E_0 \sin \frac{\alpha}{2}+4 
\int_{-\infty}^{\infty} \rho(t^{\prime})K_1(t,t^{\prime})\;d t^{\prime}\;,
\label{appen3}
\end{eqnarray}
where the kernel $K_1(t,t^{\prime})$ reads 
\begin{equation}
K_1(t,t^{\prime})=\sum_{k=-\infty}^{\infty}
\left[\frac{k}{(t-t^{\prime})^2 \tan \frac{\alpha}{2}+
(t-t^{\prime}-2k)^2 \cot \frac{\alpha}{2}}+
\frac{t^{\prime}+k}{(t+t^{\prime})^2 \tan \frac{\alpha}{2}+
(t-t^{\prime}-2k)^2 \cot \frac{\alpha}{2}}\right]\;.
\end{equation}
Taking now $E_0=1$ we arrive at (\ref{int7}).

Below we consider the asymptotic behavior of $\rho(t)$ near the 
branch points $t=0$ and $t=1$.

\textbullet\ 
$t \longrightarrow 0$.

Let us assume the power behavior $\rho(t)\propto |t|^{-2\mu_0}\cdot
\mbox{sgn}\;(t)$ and look for
the exponent $\mu_0$. The main singularity comes from integral in (\ref{int12})
in the vicinity $t^{\prime}\longrightarrow 0$. The kernel $K_2$ behaves as
\begin{eqnarray}
K_2(t,t^{\prime})\;\stackrel{t,t^{\prime}\rightarrow 0}\longrightarrow\; 
\frac{4}{\pi}\frac{t^{\prime}\tan \frac{\alpha}{2}}
{(t-t^{\prime})^2+(t+t^{\prime})^2 \tan^2 \frac{\alpha}{2}}\;,
\label{appen6}
\end{eqnarray}
that gives the asymptotic behavior
\begin{eqnarray}
\frac{\rho(t)}{Z}= 
\frac{1}{\pi}\sin \alpha \int_{-\infty}^{+\infty}  
\frac{d t^{\prime}\;t^{\prime}\;\rho(t^{\prime})}
{t^2+(t^{\prime})^2 -2 t\;t^{\prime}\cos \alpha}\;.\nonumber
\end{eqnarray}
Defining a new variable $z=t^{\prime}/t$ we obtain 
\begin{eqnarray}
\frac{\pi}{Z\sin \alpha}= \int_{0}^{+\infty}
\frac{dz\;z^{1-2\mu_0}}{1+z^2-2z\cos \alpha}+\int_{0}^{+\infty}
\frac{dz\;z^{1-2\mu_0}}{1+z^2+2z\cos \alpha}\;.
\label{appen7}
\end{eqnarray}
The evaluation of the last expression is based on the primitive fraction expansion
with further usage of standard integrals
and gives finally (\ref{int14}).

\textbullet\ 
$t \longrightarrow 1$.

Let us assume the power behavior $\rho(t)\propto |1-t|^{2\mu_1}\cdot
\mbox{sgn}\;(1-t)$ and look for the exponent $\mu_1$. It is convenient to 
define new variables $\tau=1-t,\;\tau^{\prime}=1+t^{\prime}$ and consider
the vicinity of branch point $\tau\rightarrow 0$, so $\rho(t)\propto 
|\tau|^{2\mu_1}\cdot\mbox{sgn}\;(\tau)$. In order to deal with a singular
part of the integral equation (\ref{int12}) let us differentiate the last 
over $t$
\begin{eqnarray}
\frac{2}{Z}\frac{d\;\rho(t)}{dt}=\int_{-\infty}^{+\infty}
\rho(t^{\prime}) \frac{d\;K_2(t,t^{\prime})}{dt}\;d t^{\prime}\;,\;\;\mbox{where}
\;\;\;\frac{d\;\rho(t)}{dt}\propto -2\mu_1 (1-t)^{2\mu_1-1}\;.
\label{appen10}
\end{eqnarray}
The main singularity comes from integral in (\ref{appen10}) in the vicinity 
$t^{\prime}\rightarrow 1$, or $\tau^{\prime}\rightarrow 0$, where
the kernel behaves as
\begin{eqnarray}
\frac{d K_2(t,t^{\prime})}{dt}\;\stackrel{\tau,\tau^{\prime}\rightarrow 0}
\longrightarrow\;
-\frac{4\sin \alpha}{\pi}\;\frac{\tau^{\prime}(\tau+\tau^{\prime}\cos\alpha)}
{(\tau^2+(\tau^{\prime})^2 +2 \tau\tau^{\prime}\cos \alpha)^2}\;.
\label{appen11}
\end{eqnarray}
Defining a new variable $v=\tau^{\prime}/\tau$ we obtain
\begin{eqnarray}
\frac{\pi \mu_1}{Z\sin \alpha}=\int_{0}^{+\infty}
\frac{dv\;v^{2\mu_1+1}(1+v\cos \alpha)}{(1+v^2+2v\cos \alpha)^2}+
\int_{0}^{+\infty}
\frac{dv\;v^{2\mu_1+1}(1-v\cos \alpha)}{(1+v^2-2v\cos \alpha)^2}\;.
\label{appen12}
\end{eqnarray}
Evaluating the last integrals we arrive at (\ref{int15}). 

\renewcommand{\theequation}{\thesection\arabic{equation}}
\section{Derivation of Integral Equation (\ref{int19})}
\label{appendix2}
\setcounter{equation}{0}
Reminding that the equation (\ref{int12}) is written in the sense of 
principal integral value therein, we average this equation at the large interval
$[-M,M]$, taking afterward its limit $M\rightarrow \infty$
\begin{eqnarray}
\frac{1}{2M}\int_{-M}^M\left[-\frac{2}{Z}\widetilde\rho(t)
+\frac{1}{\pi} \sin \frac{\alpha}{2}\right]\;dt=
\frac{1}{2M}\int_{-M}^M d t\int_{-M}^{+M}
\widetilde\rho(t^{\prime}) K_2(t,t^{\prime})g(t^{\prime})\;d 
t^{\prime}\;,
\label{appen14}
\end{eqnarray} 
where due to (\ref{int12a}) the left h. s. reads
\begin{eqnarray}
-\frac{2}{Z}\int_0^1\widetilde\rho(t)\;dt+\frac{1}{\pi} \sin 
\frac{\alpha}{2}
\label{appen15}
\end{eqnarray}
while the right h. s. could be simplified. Indeed, let us represent the 
kernel $K_2(t,t^{\prime})$ as follows
\begin{eqnarray}
K_2(t,t^{\prime})&=&
\frac{\left(\tan \frac{\alpha}{2}+\cot \frac{\alpha}{2}\right)\sin \pi 
(t-t^{\prime})}{\cos\pi(t-t^{\prime})-\cosh \left[\pi (t+t^{\prime})\tan 
\frac{\alpha}{2}\right]}+
\frac{\left(\tan \frac{\alpha}{2}+\cot \frac{\alpha}{2}\right)\sin \pi (t-t^{\prime})}
{\cos\pi(t-t^{\prime})-\cosh \left[\pi (t-t^{\prime})\tan \frac{\alpha}{2}\right]}+
\nonumber\\
&&\frac{1}{\pi} \cot \frac{\alpha}{2}\frac{d}{dt}
\ln \left\{\cos\pi(t-t^{\prime})-\cosh \left[\pi (t-t^{\prime})\tan 
\frac{\alpha}{2}\right]\right\}+\nonumber\\
&&\frac{1}{\pi} \cot \frac{\alpha}{2}\frac{d}{dt}
\ln \left\{\cos\pi(t-t^{\prime})-\cosh \left[\pi (t+t^{\prime})\tan
\frac{\alpha}{2}\right]\right\}\;.
\label{appen16}
\end{eqnarray}
The last three terms do not contribute to integration of the kernel over $t$,
while the first term implies
\begin{eqnarray}
-\frac{2}{Z}\int_0^1\widetilde\rho(t)\;dt+\frac{1}{\pi} \sin\frac{\alpha}{2}=
\frac{1}{2M}\int_{-M}^M \int_{-M}^M \widetilde\rho(t^{\prime}) 
K_3(t,t^{\prime})g(t^{\prime})\;d t^{\prime}d t\;,
\label{appen17}
\end{eqnarray}
where
\begin{equation}
K_3(t,t^{\prime})=\frac{2}{\sin \alpha}\cdot \frac{\sin \pi t}
{\cos\pi t-\cosh \left(\pi (2t^{\prime}+t)\tan \frac{\alpha}{2}\right)}
=-\frac{2}{\sin \alpha}\cdot{\rm Re} 
\left\{\cot \frac{\pi}{2} \left[t+i (2 t^{\prime}+t)\tan 
\frac{\alpha}{2}\right]\right\}\;.
\end{equation}
Continuing the integration of the kernel $K_3(t,t^{\prime})$ we notice that 
the function
\begin{eqnarray}
F(t^{\prime})=\lim_{M\to\infty} \int_{-M}^{+M}K_3(t,t^{\prime})\;dt
\label{appen18}
\end{eqnarray}
is 1--periodic function: $F(t^{\prime}+1)=F(t^{\prime})$, 
that follows from the structure of the kernel $K_3(t,t^{\prime})$. 
Evaluating the integral in (\ref{appen18}) we arrive at the following
\begin{equation}
F(t^{\prime})=2(1-2 t^{\prime})\;\;\mbox{for}\;\;\;0\leq t^{\prime}\leq 1\;.
\end{equation}
The final integral equation
\begin{equation}
2\left(1-\frac{1}{Z}\right)\int_0^1\widetilde\rho(t)\;dt+\frac{1}{\pi} 
\sin\frac{\alpha}{2}=
4 \int_0^1t\;\widetilde\rho(t)dt\;
\end{equation}
leads already to (\ref{int19}).

\begin{figure}[ht]
\centerline{\includegraphics*[scale=0.5]{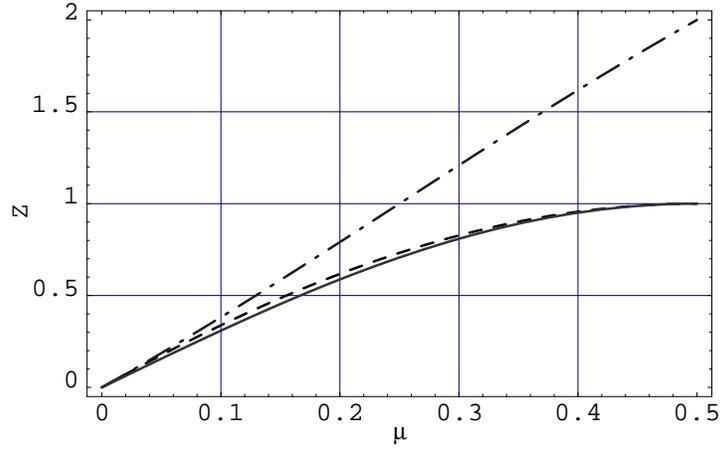}}
\caption{The branch points exponents $\mu_0(Z)$ ({\it dashed line}) and
$\mu_1(Z)$ ({\it dot-dashed line}) for rhombic unit cell with
$\alpha=\pi/3$.
The exponent $\kappa(Z)$ ({\it plain line}) for square unit cell is   
also presented.}
\label{expon1}
\end{figure}
\end{document}